\documentclass[12pt,a4paper]{article}
\usepackage{epsfig}
\def\fileversion{2.1}
\def\filedate{95/01/26}
\newlength{\dinwidth}
\newlength{\dinmargin}
\newcommand{\sinthw}{$\sin^{2}\theta_{{\mathrm w}}\space$}

\newcommand{\ee}{${\mathrm e^{+} \mathrm e^{-}}\space$}

\newcommand{\mm}{$\mu^{+}\mu^{-}\space$}

\def\today{\ifcase\month\or
 January\or February\or March\or April\or May\or June\or
 July\or August\or September\or October\or November\or December\fi
 \space\number\day, \number\year}

\newcommand{\jpg}{{\em J. Phys. G: Nucl. Part. Phys.} }   %1989 and onwards

\def\ap#1#2#3 {Ann. Phys. (NY) {\bf#1} (19#2) #3}
\def\apj#1#2#3 {Astrophys. J. {\bf#1} (19#2) #3}
\def\apjl#1#2#3 {Astrophys. J. Lett. {\bf#1} (19#2) #3}
\def\app#1#2#3 {Acta. Phys. Pol. {\bf#1} (19#2) #3}
\def\ar#1#2#3 {Ann. Rev. Nucl. Part. Sci. {\bf#1} (19#2) #3}
\def\cpc#1#2#3 {Computer Phys. Comm. {\bf#1} (19#2) #3}
\def\err#1#2#3 {{\it Erratum} {\bf#1} (19#2) #3}
\def\ib#1#2#3 {{\it ibid.} {\bf#1} (19#2) #3}
\def\jmp#1#2#3 {J. Math. Phys. {\bf#1} (19#2) #3}
\def\ijmp#1#2#3 {Int. J. Mod. Phys. {\bf#1} (19#2) #3}
\def\jetp#1#2#3 {JETP Lett. {\bf#1} (19#2) #3}
\def\jpg#1#2#3 {J. Phys. G. {\bf#1} (19#2) #3}
\def\mpl#1#2#3 {Mod. Phys. Lett. {\bf#1} (19#2) #3}
\def\nat#1#2#3 {Nature (London) {\bf#1} (19#2) #3}
\def\nc#1#2#3 {Nuovo Cim. {\bf#1} (19#2) #3}
\def\nim#1#2#3 {Nucl. Instr. Meth. {\bf#1} (19#2) #3}
\def\np#1#2#3 {Nucl. Phys. {\bf#1} (19#2) #3}
\def\pcps#1#2#3 {Proc. Cam. Phil. Soc. {\bf#1} (#2) #3}
\def\pl#1#2#3 {Phys. Lett. {\bf#1} (19#2) #3}
\def\prep#1#2#3 {Phys. Rep. {\bf#1} (19#2) #3}
\def\prev#1#2#3 {Phys. Rev. {\bf#1} (19#2) #3}
\def\prl#1#2#3 {Phys. Rev. Lett. {\bf#1} (19#2) #3}
\def\prs#1#2#3 {Proc. Roy. Soc. {\bf#1} (19#2) #3}
\def\ptp#1#2#3 {Prog. Th. Phys. {\bf#1} (19#2) #3}
\def\ps#1#2#3 {Physica Scripta {\bf#1} (19#2) #3}
\def\rmp#1#2#3 {Rev. Mod. Phys. {\bf#1} (19#2) #3}
\def\rpp#1#2#3 {Rep. Prog. Phys. {\bf#1} (19#2) #3}
\def\sjnp#1#2#3 {Sov. J. Nucl. Phys. {\bf#1} (19#2) #3}
\def\spj#1#2#3 {Sov. Phys. JEPT {\bf#1} (19#2) #3}
\def\spu#1#2#3 {Sov. Phys.-Usp. {\bf#1} (19#2) #3}
\def\zp#1#2#3 {Zeit. Phys. {\bf#1} (19#2) #3}
\def\sci#1#2#3 {Science {\bf#1} (19#2) #3}
\begin{document}

\begin{titlepage}
\begin{flushright}
{ETHZ-IPP  PR-96-01} \\
21 March, 1996 \\
\end{flushright}
 
\vspace{2.5cm}
\bigskip
\begin{center}
{\Large \bf 
Neutral Current Interference in the TeV Region;}
\end{center}
\begin{center}
{\Large \bf the Experimental Sensitivity at the LHC}
\end{center}
\smallskip
\smallskip
\bigskip

\begin{center}
{\large Michael Dittmar}
\end{center}

\begin{center}
Institute for Particle Physics (IPP), ETH Z\"{u}rich, \\
CH-8093 Z\"{u}rich, Switzerland
\end{center}

\begin{center}
{\large \bf Abstract} 
\end{center}
The possibilities to measure lepton 
forward--backward asymmetries 
at the LHC in the reaction 
pp($q_{i}\overline{q_{i}})\rightarrow \ell^{+} \ell^{-}$
are studied for dilepton events with 
masses above 400 GeV.   
It is shown that such measurements 
allow accurate tests of the neutral current 
interference structure up to about 2 TeV center of mass energies. 
The sensitivity of asymmetries at the LHC 
to new physics is demonstrated within 
the context of quark compositeness and exotic Z$^{'}$ scenarios.
\bigskip
\bigskip
\bigskip
\bigskip
%\begin{center}
%{\large submitted to Zeit. Phys. C}
%\end{center}
\bigskip
\end{titlepage}
% \newpage
\section{Introduction}
Measurements of forward--backward charge asymmetries, 
$\mathrm{A_{fb}}$, are traditionally a domain of \ee\ collider 
experiments. Evidence for the $\gamma$--Z interference came 
from the observation of the non zero value of $\mathrm{A_{fb}}$
in the reaction \ee\ $\rightarrow$ \mm\ 
at PETRA around 1982/1983 \cite{bib-afbpetra}.
Furthermore, $\mathrm{A_{fb}}$ measurements 
with quarks and leptons at the Z peak provide 
precise \sinthw\ determinations \cite{bib-lepz0}.
The sensitivity of $\mathrm{A_{fb}}$ measurements 
to new physics like additional Z$^{'}$ bosons or compositeness
has been shown in many details for future 
high energy \ee\ collider experiments
\cite{bib-eefuture}. 
Lepton asymmetries in the reaction 
p--$\bar{\mathrm p} \rightarrow \mathrm{e}^{+} \mathrm{e}^{-}
(\mu^{+} \mu^{-})$ can be defined 
with respect to the proton direction. Such  
asymmetry measurements, \cite{bib-ppbar},
are currently less precise than the corresponding \ee\ 
results and have so far only been performed close to 
the Z resonance.
\par
High mass dilepton events in p--p collisions 
originate from the annihilation
of valence quarks with sea--antiquarks or  
from the annihilation of sea--quarks with sea--antiquarks.
As the valence quarks have on average a much larger momentum 
than the sea--antiquarks, the boost direction 
of the dilepton system approximates the quark direction.
A lepton asymmetry can thus be expected with respect to the 
boost direction. In contrast, dilepton 
events which originate from the annihilation 
of quark pairs from the sea must be symmetric. 
\par
The sensitivity of $\mathrm{A_{fb}}$
measurements to Z$^{'}$ bosons 
at future multi TeV p--p and p--$\bar{\mathrm p}$
has been discussed by several Authors \cite{bib-ppasym}. 
In particular P. Langacker et al.
have studied the sensitivity of on--resonance asymmetries
to the couplings of hypothetical
new bosons at future multi TeV hadron colliders.
Such on--resonance asymmetry measurements at the LHC 
were simulated in detail for the 1990 workshop in 
Aachen \cite{bib-lhcaachen} and 
also for the design studies of ATLAS \cite{bib-atlastp}
and CMS \cite{bib-cmstp}. The possibility of   
an asymmetry measurement at the Z peak
with a dedicated LHC experiment 
has also been discussed \cite{bib-lhcz0}.
\par
However, asymmetries from 
continuum dilepton events at very high energy pp colliders 
as an additional tool to study the neutral current interference
have so far only been studied theoretically.  
J. Rosner has investigated the sensitivity of 
lepton asymmetries with on--resonance, off--resonance and 
continuum dilepton events at the SSC. 
He concluded that the combination of all these 
measurements should provide the best sensitivity to 
new physics in neutral current reactions 
\cite{bib-rosner87}.
\par   
With the more clearly defined experimental capabilites
of the LHC experiments ATLAS \cite{bib-atlastp} and CMS
\cite{bib-cmstp} and an expected integrated yearly luminosity 
of up to 100 fb$^{-1}$ per experiment 
it becomes interesting to perform a 
simulation of such an asymmetry measurement at the LHC.
Such an analysis and its sensitivity to new physics
is described in the following. 
   
\section{Asymmetries in reactions with quark and 
lepton pairs}

The reactions 
pp($q_{i} \overline{q_{i}}) \rightarrow e^{+} e^{-} $ and  
pp($q_{i} \overline{q_{i}}) \rightarrow \mu^{+} \mu^{-} $,
as well as the inverse reaction 
$e^{+} e^{-} \rightarrow q_{i} \overline{q_{i}} $  
are described by the exchange of neutral vector bosons.
Within the Standard Model the couplings 
of the photon and the Z to quarks
and leptons are known and precise calculations for 
the interference between the photon and the Z 
exist. The $\gamma$--Z interference results in large 
forward--backward asymmetries for 
center of mass energies well above the Z peak.
For center of mass energies above 250~GeV, 
essentially constant asymmetries of about
61\% are expected for the above reactions with up--type  
and down--type quarks.
New phenomena in the TeV range, 
like Z$^{'}$ bosons \cite{bib-ppasym}
or contact interactions 
between quarks and lepton due to compositeness \cite{bib-compo}
might considerably modify this picture.
Fermion asymmetry measurements at high center of 
mass energies are consequently an  
excellent tool to search for and perhaps study such new phenomena. 
 
Asymmetries with quarks and leptons in the final state
up to $m_{\ell \ell} (=\sqrt{s}) \approx$ 200 GeV 
will be measured in detail at LEP II and 
the region of perhaps up to $m_{\ell \ell} \approx$ 500 GeV 
will be investigated at the upgraded TEVATRON p$\bar{\mathrm p}$ 
collider \cite{bib-rosner95}. To exploit the 
higher energies at the LHC, we concentrate in this 
study on high mass dilepton events ($m_{\ell \ell} > 400$ GeV).  

In pp collisions, unlike in \ee\ collider experiments,
the center of mass frame is different from the laboratory frame. 
However, the four momenta of the 
dilepton system are measured and one can calculate 
the lepton (electron or muon) angle $\theta^{*}$ 
in the dilepton center of mass frame.
The lepton asymmetry $\mathrm{A_{FB}^{\ell}}$ is defined 
from the angular distribution $\cos \theta^{*}$ with respect to the 
quark direction and can be obtained best 
with an unbinned maximum likelihood 
fit to the cos $\theta^{*}$ distribution given by:
\begin{eqnarray}
\frac{ d \sigma } { d\cos \theta^{*}}
\propto 3/8 (1 + \cos^{2} \theta^{*}) + {\mathrm A_{FB}^{\ell}} \cos \theta^{*} 
\end{eqnarray}

To simulate an asymmetry measurement with dilepton events
at the LHC (with 14 TeV pp collisions) 
the PYTHIA Monte Carlo \cite{bib-pythia57} is used. 
The PYTHIA dilepton event generators 
allow to simulate the Standard Model predictions with
a photon and a Z exchange as well as modifications 
due to compositeness \cite{bib-pythiagzcomp} or the 
possibilities of the additional 
exchange of a Z$^{'}$ including the interference between
$\gamma$, Z and Z$^{'}$ \cite{bib-pythiagzzp}.

For the Standard Model simulation of dilepton events
with a mass larger than 400 GeV, one finds that 
about 70\% originate from the annihilation of
$u\bar{u}$ quarks, 21\% from $d\bar{d}$ quarks and about
9\% are from the annihilation of $s\bar{s}$,
$c\bar{c}$ and $b\bar{b}$ sea--quarks. 
Uncertainties from the quark flavour composition
are not important for comparisons of a measurement with the 
Standard Model as essentially identical lepton asymmetries 
are predicted for the annihilation of $u\bar{u}$ and $d\bar{d}$ 
quark pairs and dilepton masses above 400 GeV.
However, at a pp collider, 
observable asymmetries must come from those dilepton 
events which originate from the annihilation of a valence quark 
(u and d quarks) and the corresponding sea--antiquark. 
The fraction of events which will show 
measurable asymmetries depends therefore on the mass
and on the boost of the dilepton system.
 
As a first step of the simulation, the charge asymmetries 
are determined with respect to the quark direction taken
from the generator. Using this approach
lepton asymmetries of $\approx$ 61\% 
are obtained from the fit to the 
$\cos \theta^{*}$ distribution of dilepton events and the 
Standard Model simulation. The asymmetry is essentially 
independent of the mass and rapidity of the lepton system.
If instead a random assignment for the 
quark direction is used, the angular distribution of the leptons  
is well described by a $1 + \cos^{2} \theta^{*}$ function 
with an asymmetry of zero.

As the original quark direction is not known in pp collisions one
has to extract it from the kinematics of the dilepton system. 
For this analysis, the quark direction is taken from 
the boost direction of the dilepton system 
with respect to the beam axis (the z--axis).
The correctness of this assignment is studied 
as a function of the dilepton rapidity
$\mathrm{y} = 1/2 \times ln [(E + p_{z})/(E - p_{z})]$. 
The rapidity distribution for dilepton events with masses
above 400 GeV is shown in Figure 1a for all dilepton events 
and for the subsample of events where 
the sign of the boost direction and the 
quark direction agree. As shown in Figure 1b, 
the fraction of events with a correctly assigned 
quark direction increases as a function of the rapidity. 
For small rapidities ($|\mathrm{y}| < 0.2$) 
essentially all dilepton events
originate from the annihilation of sea--quarks with sea--antiquarks
and an almost random assignment is made for the quark direction. 
The fraction of events with a correctly assigned quark direction  
increases to more than 80\% at a rapidity of one.

\begin{figure}[htb]
\begin{center}\mbox{
\epsfig{file=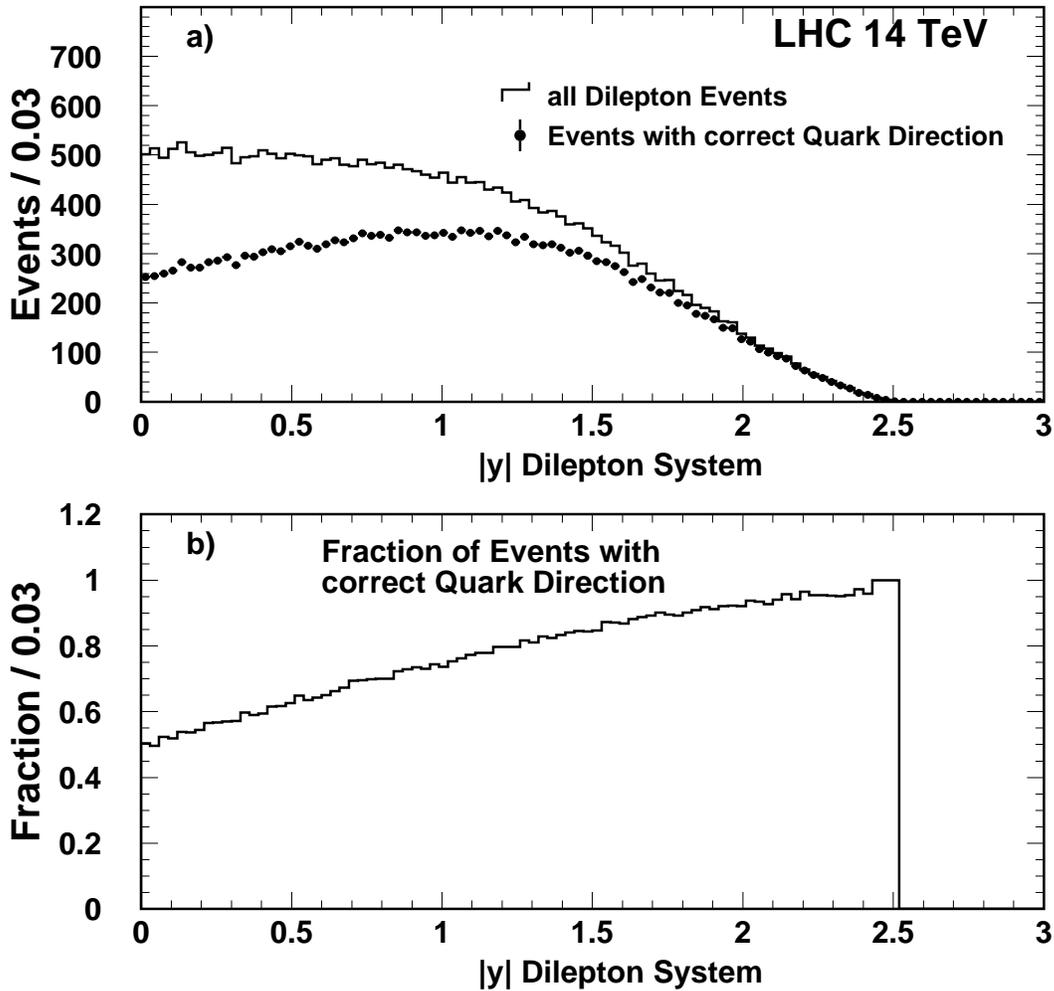,height=15 cm,width=15cm}}
\end{center}

\caption[fig1ab]{Rapidity $| \mathrm{y} |$ 
distribution of the dilepton  
system with respect to the beam direction for all events and the
subsample where the sign of the rapidity agrees with the  
quark direction (a). The fraction of events with
a correctly assigned quark direction is shown in (b).
}
\end{figure}
\clearpage

Uncertainties for the quark direction assignment 
and from the fraction of events which originate from
the annihilation of sea--quark pairs
have been studied with a variety of structure functions
\cite{bib-strucfunc} as implemented in PYTHIA.
From this study we find that for 
dilepton masses above 400 GeV these
uncertainties are essentially negligible compared to  
the achievable statistical precisions of 
the $\mathrm{A_{fb}^{\ell}}$ measurements.
However, for dilepton masses below 200 GeV, the uncertainties
are more important for asymmetry measurements
and require special attention.
As this center of mass energy range will already be covered 
by LEP II, we concentrate in the following only on
dilepton events with high mass.

\section{Simulation of an Asymmetry Measurement}

According to the design characteristics of ATLAS \cite{bib-atlastp} 
and CMS \cite{bib-cmstp} high identification 
efficiency for isolated electrons and muons with
excellent energy and momentum resolutions 
are expected down to pseudorapidities of $|\eta | < 2.5 $
and for the highest LHC 
luminosities (10$^{34}$cm$^{-2}$sec$^{-1}$). 
For example, using the beam spot constraint, the momentum 
resolution for 1 TeV muons and electrons should be 
better than 10\% and no charge confusion is expected up to 
much higher momenta. With this accuracy,
dilepton mass resolutions of better than 5--10\% for dimuon events
and about 1--2\% for dielectron events 
can be expected for masses of about 1 TeV.
\par
For this study, PYTHIA 
events of the type pp $\rightarrow e^{+} e^{-} (\gamma)$ and 
pp $\rightarrow \mu^{+} \mu^{-} (\gamma)$ are 
simulated at a center of mass energy of 14 TeV. The produced leptons 
are in general isolated and are essentially back to back in the 
plane transverse to the beam direction. 
Dilepton events, either 
dielectron or dimuon events, are accepted 
if the following criteria are fulfilled:
\begin{itemize}
\item The minimum $p_{t}$ of each charged lepton 
should be larger than 20 GeV.
\item
The pseudorapidity $|\eta|$ of each lepton 
should be smaller than 2.5.
\item
The two leptons must have opposite charge.
\item
The two leptons should be back to back
in the plane transverse to the beam direction with an 
opening angle between the two leptons of more than 160$^{\circ}$.
\item
The dilepton mass has to be larger than 400 GeV. 
\end{itemize}

Especially for dilepton pairs with masses in the 
TeV region these criteria seem to be sufficient
to select essentially background free events.   
However, if required by backgrounds,
isolation criteria for electrons and muons and  
veto criteria against events which have a 
large jet activity, can be applied without a significant loss
in statistics. 

For an integrated luminosity of 100~fb$^{-1}$
about 28k dilepton events with a mass above 400~GeV are 
accepted with these criteria. Table 1 shows the 
number of expected events for different mass intervals.
For a real experiment the number of events will be further reduced
by the imperfect geometrical detector coverage and lepton detection 
efficiencies $\epsilon$. The number of events, given in Table 1 
should be multiplied by $\epsilon^{2}$ and the 
estimated asymmetry errors will increase by roughly 
$1/\epsilon$. Accurate efficiency estimates are difficult to make,
but values of 90\% and more have been used for 
other simulations \cite{bib-atlashiggs}.

As discussed in section 2, the lepton asymmetry 
is obtained from the $\cos \theta^{*}$ distribution in the dilepton 
center of mass frame. The $\cos \theta^{*}$ distribution 
for dilepton masses between 0.75 TeV and 1.25 TeV 
and an absolute rapidity of more than 0.8 is shown in Figure 2a 
before and after the selection criteria are applied.
As expected, the used lepton selection criteria
effect only large $|\cos \theta^{*}|$ values.
\begin{figure}[htb]
\begin{center}\mbox{
\epsfig{file=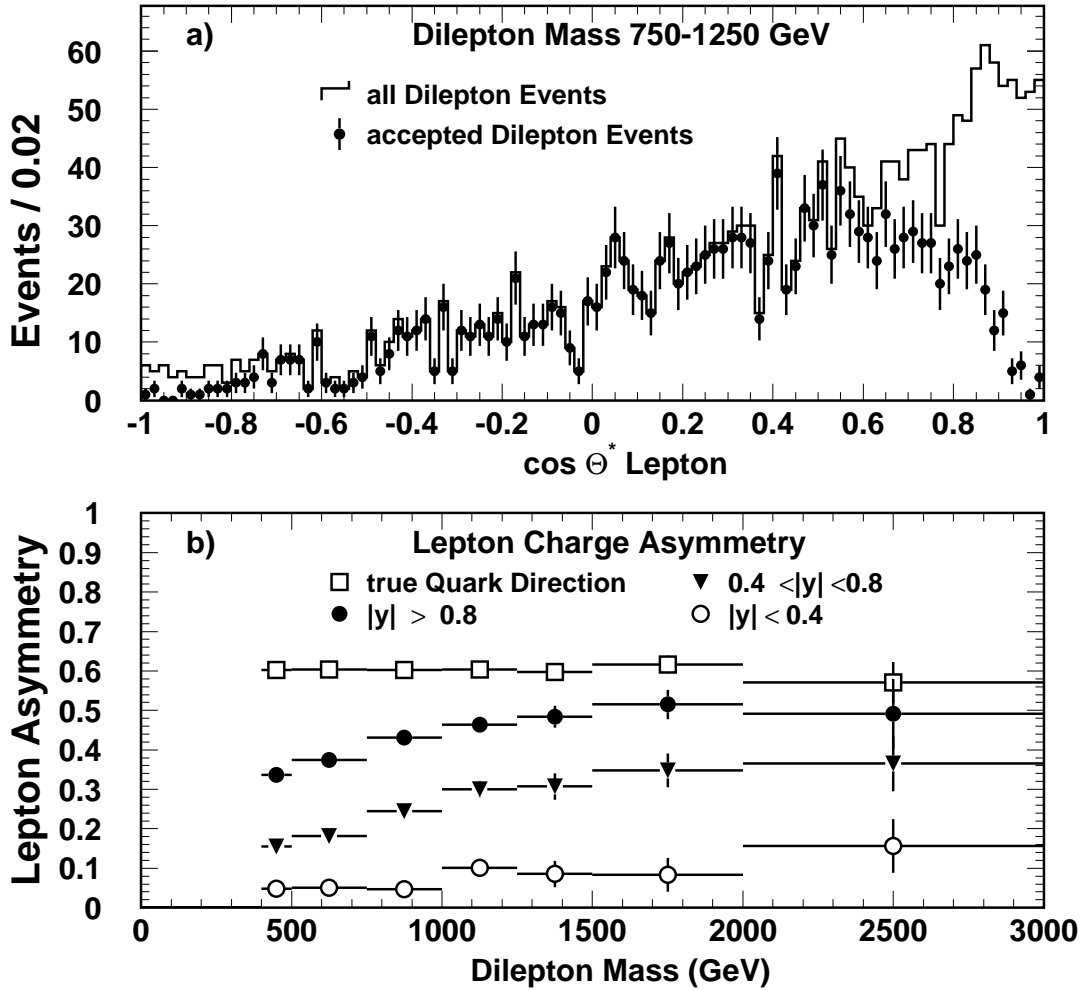,height=15 cm,width=15cm}}
\end{center}
\caption[fig2ab]{(a) The Standard Model predictions for  
the $\cos \theta^{*}$ distribution of the 
negatively charged lepton with and without acceptance criteria
for a mass of the dilepton system in the 
interval 0.75--1.25 TeV and an 
absolute rapidity of the dilepton system larger than 0.8. 
Asymmetries for different dilepton mass  
and rapidity intervals are shown in   
(b).
}
\end{figure}
\clearpage  

Lepton asymmetries are now determined 
from the $\cos \theta^{*}$ distribution 
of the negatively charged lepton 
with respect to the boost direction of the dilepton 
system along the z axis.
For various dilepton mass and rapidity intervals, 
the $\cos \theta^{*}$
distribution is fitted using the unbinned maximum likelihood method.
These lepton asymmetries are shown in Figure 2b as a function of the 
dilepton mass (using about 10 times the 
integrated luminosity expected per experiment). 
Large and statistically significant 
asymmetries are found for large dilepton rapidities
up to masses of about 2 TeV. 

The extracted asymmetries are smaller than 
the ones expected from the Standard Model and depend on the 
mass and rapidity of the dilepton system.  
This difference comes from two sources, the contributions from 
events which originate from the annihilation of sea--quarks
which must have zero asymmetry and from 
the cases where the valence quark carries a smaller momentum
than the sea--antiquark. Both contributions reduce the 
asymmetries and can be considered as random background. 
This background is especially important for low dilepton masses. 
Possible uncertainties from these backgrounds 
were studied and found to be much smaller 
than the obtainable statistical errors 
from an integrated luminosity of 100~fb$^{-1}$.
The expected Standard Model asymmetries, uncorrected for 
backgrounds, with statistical errors 
assuming an integrated luminosity of 100~fb$^{-1}$
are given in Table 1. 
\begin{table}[htb]
\vspace{0.3cm}
\begin{center}
\begin{tabular}{|c|c|c|c|c|}
\hline
mass($\ell^{+}\ell^{-}$)& Number of Events &
\multicolumn{3}{|c|}{$ \mathrm{A_{FB}^{\ell}}$ } \\
\hline
[TeV] & (for 100 fb$^{-1}$) & 
$|y|<0.4$ & $0.4 < |y| < 0.8$ & $|y|>0.8$ \\ \hline
0.40-0.50 & 14100 & 0.05$\pm$0.02 & 0.16$\pm$0.02 & 0.34$\pm$0.01\\
0.50-0.75 & 10300 & 0.05$\pm$0.02 & 0.18$\pm$0.02 & 0.37$\pm$0.01\\
0.75-1.00 & ~2270 & 0.05$\pm$0.04 & 0.24$\pm$0.04 & 0.43$\pm$0.03\\
1.00-1.25 & ~~723 & 0.10$\pm$0.06 & 0.30$\pm$0.06 & 0.46$\pm$0.05\\
1.25-1.50 & ~~261 & 0.09$\pm$0.10 & 0.31$\pm$0.10 & 0.48$\pm$0.08\\
1.50-2.00 & ~~155 & 0.08$\pm$0.12 & 0.35$\pm$0.13 & 0.52$\pm$0.12\\
$>$2.00   & ~~~42 & 0.16$\pm$0.21 & 0.37$\pm$0.21 & 0.49$\pm$0.28\\
\hline
\end{tabular}\vspace{0.3cm}
\end{center}
\caption{The Standard Model predictions for the number of
accepted dilepton events, assuming 100\% lepton identification
for the used geometrical selection with
an integrated luminosity of 100~fb$^{-1}$ and different mass
bins. The expected asymmetries with their statistical errors
for the different mass and rapidity intervals 
are also given.}
\end{table}

It is interesting to compare the estimated  
accuracy of asymmetry measurements at the LHC
with measurable quark asymmetries 
at a high energy linear \ee\ collider. 
The most accurate and efficient 
asymmetry measurements with quarks in the final 
state can be obtained  
for the reaction $e^{+} e^{-} \rightarrow b \overline{b}$. 
To obtain asymmetry errors at $m_{\ell \ell} \approx 1$~TeV 
of about 2--3\%, at least 1000 accepted and tagged b--events 
are required. The cross section for the reaction 
$e^{+} e^{-} \rightarrow b \overline{b}$ at 1~TeV center of mass
energy is approximately 100~fb. 
Assuming an optimistic 
efficiency of about 10\% to identify b events and measure
correctly the charge of b flavoured jets, one can
estimate the required LHC equivalent yearly 
luminosity to be about 100~fb$^{-1}$,
corresponding to an average luminosity of 
$10^{34}$sec$^{-1}$cm$^{-2}$. 
 
\section{$\mathbf{\mathrm{A_{FB}^{\ell}}}$ and Exotica}

We now discuss the sensitivity of the $\mathrm{A_{FB}^{\ell}}$ 
measurements at the LHC with respect to  
two different exotic physics scenarios as implemented in PYTHIA. 
These are (a) composite models according to the model 
by E. Eichten et al.
\cite{bib-compo} and (b) a heavy 
Z$^{'}$ with quark and lepton couplings  
identical to the Standard Model Z \cite{bib-pythiagzzp}.
To demonstrate the sensitivity of asymmetry measurements two
extreme scenarios for the 
total Z$^{'}$ width have been used. 
The width has been modified by changing the decay rate
Z$^{'}\rightarrow$W$^{+}W^{-}$ to very large values. 
For various other Z$^{'}$ scenarios with their detailed 
lepton asymmetry predictions we refer to the 
literature \cite{bib-zprime}. 
The different model predictions for dilepton event rates
are determined with the above selection criteria. For the 
asymmetries it is further required that  
the dilepton rapidities $| \mathrm{y} | $ are larger than 0.5.

The compositeness scenario is simulated such 
that u and d quarks are composite with a scale 
$\Lambda^{\pm}$, the sign indicates 
either a positive or a negative interference term.
As a result of the contact interaction one expects 
a flattening of the steeply falling dilepton mass distribution
as shown in Figure 3a for 
$\Lambda^{\pm}$ of 30 TeV and 
for the Standard Model ($\Lambda^{\pm}=\infty$). Detailed
discussions of these cross section changes 
can be found in \cite{bib-compo}.   
In Table 2, dilepton event rates and asymmetries 
are given for different masses and $\Lambda^{\pm}$ scales.
The central values for the different models 
have been obtained from a high 
statistic simulation equivalent to about 1000 fb$^{-1}$.
\begin{table}[htb]
\vspace{0.3cm}
\begin{center}
\begin{tabular}{|c|c|c|c|c|c|}
\hline
mass($\ell^{+}\ell^{-}$) 
& \multicolumn{5}{|c|}{number of events (100 fb$^{-1}$)} \\ \hline
[TeV]& $\Lambda^{\infty}$ & $\Lambda^{+}$=20 TeV& $\Lambda^{+}$=30 TeV & 
$\Lambda^{-}=20$ TeV & $\Lambda^{-}$=30 TeV
\\ \hline
0.4-0.5 & 14100 & 13940&14020 & 14380&14118 \\
0.5-0.6 & ~6200 & ~6127&~6137 & ~6462&~6351 \\
0.6-0.9 & ~5755 & ~5591&~5655 & ~6180&~5923 \\
0.9-1.1 & ~~978 & ~~885&~~940 & ~1152&~1023 \\
1.1-1.4 & ~~525 & ~~466&~~487 & ~~674&~~559 \\
1.4-3.0 & ~~267 & ~~320&~~257 & ~~558&~~353 \\
\hline
& \multicolumn{5}{|c|}{$ \mathrm{A_{FB}^{\ell}} (| y | > 0.5)$ } \\
\hline
0.4-0.5 &0.283$\pm$0.010 &0.270 &0.279 & 0.291&0.288\\
0.5-0.6 &0.320$\pm$0.015 &0.296 &0.310 & 0.323&0.324\\
0.6-0.9 &0.345$\pm$0.015 &0.325 &0.344 & 0.373&0.359\\
0.9-1.1 &0.392$\pm$0.035 &0.342 &0.356 & 0.434&0.416\\
1.1-1.4 &0.421$\pm$0.048 &0.321 &0.368 & 0.498&0.465\\
1.4-3.0 &0.451$\pm$0.068 &0.345 &0.383 & 0.554&0.484\\
\hline
\end{tabular}\vspace{0.3cm}
\end{center}
\caption{Accepted number of dilepton events and expected 
asymmetries for $|\mathrm{y}|$ larger than 0.5
for different compositeness scales $\Lambda^{\pm}$, 
corresponding to an integrated luminosity 
of 100~fb$^{-1}$. For the Standard Model ($\Lambda^{\infty}$) the  
statistical errors of the asymmetry results are also given.
}
\end{table}
Taking dilepton cross section uncertainties of 
up to 20\% into account,  
the event rate with masses above 1 TeV
alone will perhaps not be significant enough to observe 
compositeness with $\Lambda^{\pm}$ scales
of about 25--30 TeV.
However, the asymmetries, as shown in Figure 3b
and Table 2 will improve the sensitivity 
to compositeness phenomena considerable.
The combination of cross section and asymmetry measurements
should thus allow to increase the sensitivity at the two sigma 
level to $\Lambda^{\pm}$ values of almost 30 TeV.
\begin{figure}[ht]
\begin{center}\mbox{
\epsfig{file=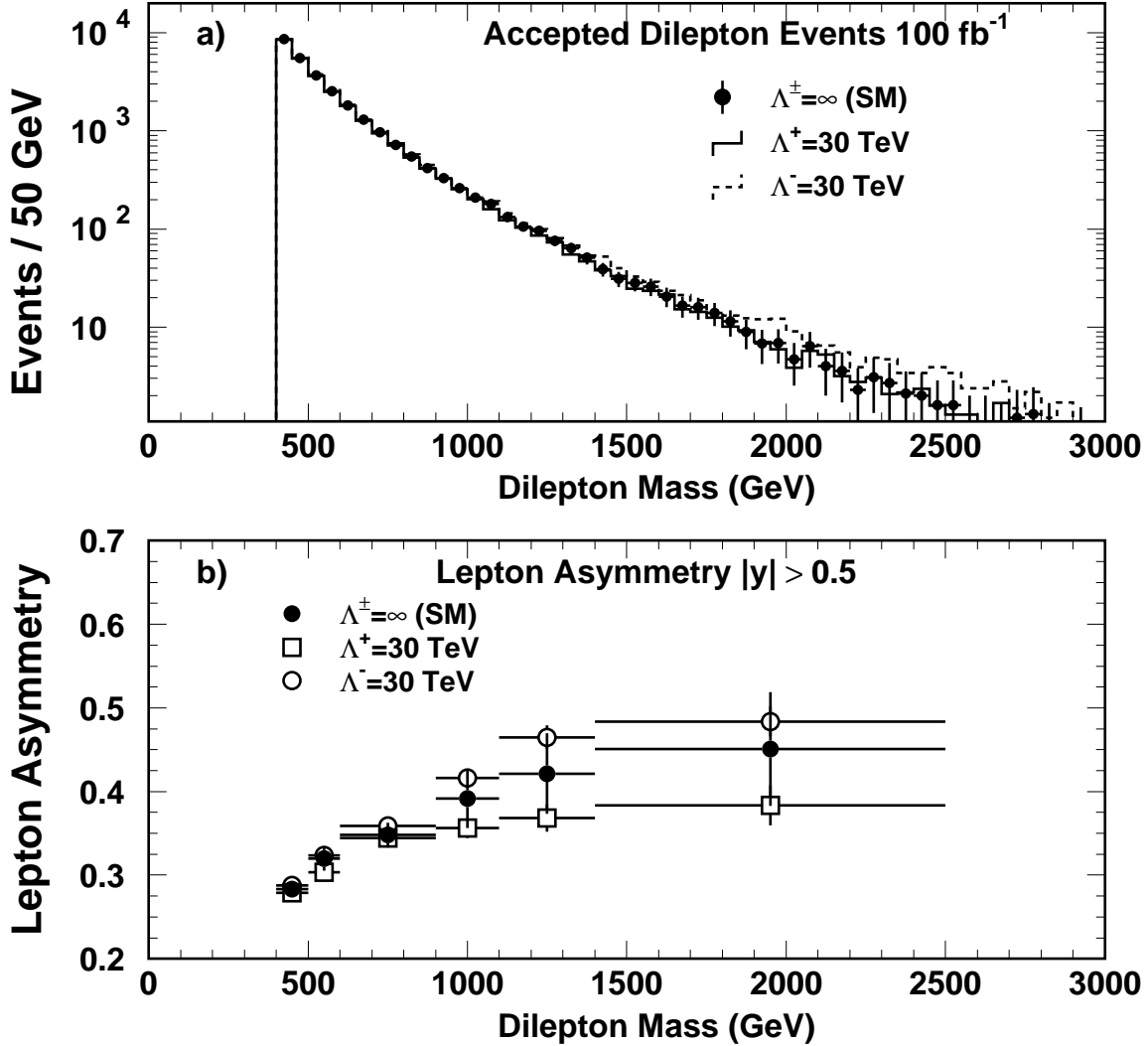,height=16 cm,width=16cm}}
\end{center}
\caption[fig3ad] {Expected dilepton mass distributions (a)
and lepton asymmetries (b) 
for the Standard Model and for quark compositeness with 
different $\Lambda^{\pm}$ values.
}
\end{figure}
% \clearpage
  
The sensitivity of off-- and on--resonance  
lepton asymmetry measurements to a Z$^{'}$ is demonstrated with  
two extreme scenarios. For this study a Z$^{'}$ with a 
mass of 1 TeV and 
a width of either $\approx$120 GeV or 
of $\approx$860 GeV are used. 
The resulting dilepton mass distributions from the  
exchange of photon, Z and Z$^{'}$ 
are compared in Figure 4a with the ones from the 
Standard Model with photon and Z exchange only.
Clear mass peaks will demonstrate 
the presence of such a Z$^{'}$ boson up to a few TeV.
Only for a very broad Z$^{'}$ the dilepton 
cross section measurement alone might not be sufficient.
However, lepton asymmetries even far away from the Z$^{'}$ peak,
as shown in Figures 4b, reveal large deviations from the 
Standard Model scenarios. We have also studied the 
continuum asymmetries for higher Z$^{'}$ masses with 
relatively large width
and find significant deviations from the Standard Model 
neutral current structure up to masses of about 2.5 TeV. 
\begin{figure}[ht]
\begin{center}\mbox{
\epsfig{file=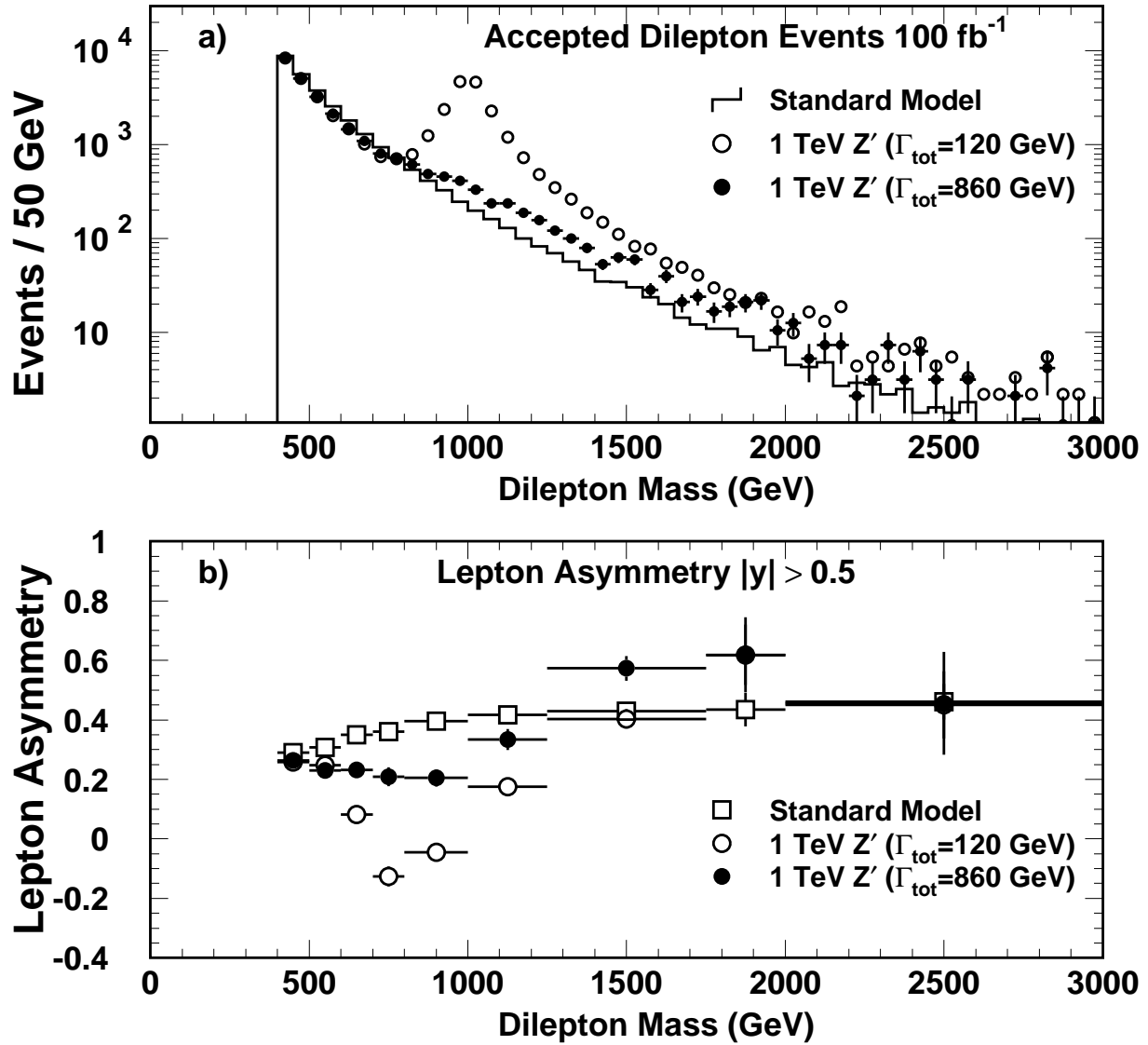,height=17 cm,width=17cm}}
\end{center}
\caption[fig4ab]{a) Expected dilepton mass distributions (a) 
and asymmetries (b) 
for the Standard Model and for two exotic Z$^{'}$ scenarios.
}
\end{figure}
\clearpage
  
\section{Summary}

We have demonstrated the feasibility of 
lepton forward--backward asymmetry measurements
with dilepton continuum events at the LHC. 
Such measurements, assuming the expected performance 
of ATLAS and CMS, appear to be almost straight forward. 
For an integrated luminosity of 100 fb$^{-1}$, lepton 
asymmetry measurements will allow an accurate study of 
neutral current interference effects 
up to dilepton masses of about 2 TeV.
The obtainable asymmetry accuracy is comparable with the ones at 
a 1 TeV linear \ee\ collider with
quark final states and an integrated luminosity of about 100 fb$^{-1}$.

From a study of two different exotic scenarios, 
quark compositeness or exotic Z$^{'}$ models, 
we conclude that continuum lepton asymmetries at the LHC
can be considered as an additional and accurate 
experimental tool to observe and study new physics in the TeV 
center of mass energy domain.   

%-----------------------------------------------------------------
\vspace{2.cm}

{\bf \large Acknowledgements}

{\small

I am grateful to E. Richter--Was for many detailed 
discussions about this analysis and her critical 
comments on the manuscript. 
I also would like to thank F. Pauss for several stimulating 
discussions and F. Behner for his suggestions on how to  
write an efficient fitting program.}


\begin{thebibliography}{99}
\bibitem{bib-pythiagzcomp} The PYTHIA subprocesses 1 and 165 are used
to simulate dilepton events according to 
the formalism given in \cite{bib-compo}.
\bibitem{bib-strucfunc} The CTEQ2 structure functions 
(mainly CTEQ2L, the PYTHIA 5.7 default) 
have been used for the simulation.
\end{thebibliography}
\end{document}